\documentclass[aps,prl,twocolumn,showpacs,superscriptaddress]{revtex4-1}

\usepackage{graphicx}
\graphicspath{{Images/}}
\usepackage[utf8]{inputenc}   
\usepackage{dcolumn}
\usepackage{color}
\usepackage{xcolor}
\usepackage{bm}
\usepackage{hyperref}

\usepackage{amsmath}

\begin{document}

\preprint{APS/123-QED}

\title{Pinning dependent field driven domain wall dynamics and thermal scaling in an ultrathin Pt/Co/Pt magnetic film}

\author{J. Gorchon}%
\affiliation{Laboratoire de Physique des Solides, Universit\'{e} Paris-Sud, CNRS, UMR8502, 91405 Orsay, France}

\author{S. Bustingorry}
\affiliation{Centro At\'{o}mico Bariloche, Comisi\'{o}n Nacional de Energ\'{i}a At\'{o}mica, Bariloche, R\'{i}o Negro, Argentina}

\author{J. Ferr\'{e}}%
\affiliation{Laboratoire de Physique des Solides, Universit\'{e} Paris-Sud, CNRS, UMR8502, 91405 Orsay, France}

\author{V. Jeudy}%
\email{vincent.jeudy@u-psud.fr}
\affiliation{Laboratoire de Physique des Solides, Universit\'{e} Paris-Sud, CNRS, UMR8502, 91405 Orsay, France}
\affiliation{Universit\'{e} Cergy-Pontoise, 95000 Cergy-Pontoise, France}

\author{A. B. Kolton}
\affiliation{Centro At\'{o}mico Bariloche, Comisi\'{o}n Nacional de Energ\'{i}a At\'{o}mica, Bariloche, R\'{i}o Negro, Argentina}

\author{T. Giamarchi}
\affiliation{DPMC, University of Geneva, 24 Quai Ernest Ansermet, 1211 Geneva, Switzerland}

\date{\today}

\begin{abstract}
Magnetic field-driven domain wall motion in an ultrathin Pt/Co(0.45nm)/Pt ferromagnetic film with perpendicular anisotropy is studied over a wide temperature range. Three different pinning dependent dynamical regimes are clearly identified: the creep, the thermally assisted flux flow and the depinning, as well as their corresponding crossovers. The wall elastic energy and microscopic parameters characterizing the pinning are determined. Both the extracted thermal rounding exponent at the depinning transition, $\psi=$0.15, and the Larkin length crossover exponent, $\phi=$0.24, fit well with the numerical predictions. 
\end{abstract}

\maketitle


Many areas in physics~\cite{blatter,3,4,kleemann,1,2,5,6} such as magnetic and ferroelectric domain walls motion, contact lines in wetting, crack propagation, vortex lines motion in type II superconductors..., involve the displacement of elastic object or interface in a weakly disordered medium.
How the velocity of motion depends on the driving force $f$ poses important fundamental questions~\cite{1,2,5,6}.
In the absence of disorder or for a large $f$, motion is limited by dissipation and the interface moves in a flow regime, with a velocity essentially proportional to $f$. However in real materials the presence of disorder leads to pinning which dramatically modifies the response to the force. At zero temperature this leads to the existence of a depinning force $f_{dep}$, below which no motion takes place. At finite temperature $T$ the combination of the applied force, collective pinning and thermal effects leads to an extremely rich dynamical behavior, which has been the focuss of many theoretical~\cite{1,2,5,6,ferrero-CRAS} and experimental studies~\cite{7,8,tybell,paruch,kim-larkin,ferre-CRAS,paruch-CRAS}.

On the experimental front the controlled investigation of this dynamics is very difficult and Pt/Co/Pt ultrathin ferromagnetic films with perpendicular anisotropy proved to be an archetypal 2D-disordered system to test theory \cite{7,8,kim-larkin,9,caysoll,10}. 
In this system the force $f$ is the applied field $H$, and domain walls (DWs) mimic elastic interfaces. Measurements as a function of the field, allow to unambiguously evidence the very non linear response $\ln v(H) \sim H^{-\mu}$ expected at small fields, the so called \textit{creep} regime \cite{7,8,9}. They confirm the predicted value $\mu=$1/4 of the exponent and its relation with the exponent
measuring the roughness of the interface at equilibrium \cite{1,4,blatter,7}.
Measurements as a function of the temperature \cite{12,13} for small fields,
confirm the role played by thermal activation over barriers by obtaining a prefactor in the exponential varying as $1/T$.

At larger fields, and in particular close to the depinning transition the situation, both theoretically and experimentally, is much less clear. At the depinning field $H_{dep}$ the response was predicted to follow a power law behavior, $v(H_{dep}) \sim T^\psi$, where $\psi$ is the thermal rounding exponent\cite{5}.
This behavior was checked indirectly \cite{11} in experiments with a constant $T$ but with Pt/Co/Pt layers of different anisotropy. However a true temperature dependent analysis close to depinning is still lacking, as well as bridging the gap between the very low field \textit{creep} regime $H \ll H_{dep}$ and the depinning one
$H \sim H_{dep}$. For this, experiments with a full range of magnetic fields and various temperature is necessary.

In this letter, we perform such a study by exploring in a single Pt/Co/Pt ultrathin layer the DW dynamics between 50 and 300K. We evidence three distinct pinning dependent regimes and determine the corresponding crossover fields. We provide a consistent theoretical description
of the full dynamical range, allowing in addition, through the testing of the predicted universal
scalings, to obtain accurately the (non-universal) microscopic quantities controlling DW dynamics.

\begin{figure}
\includegraphics[width=.9\columnwidth]{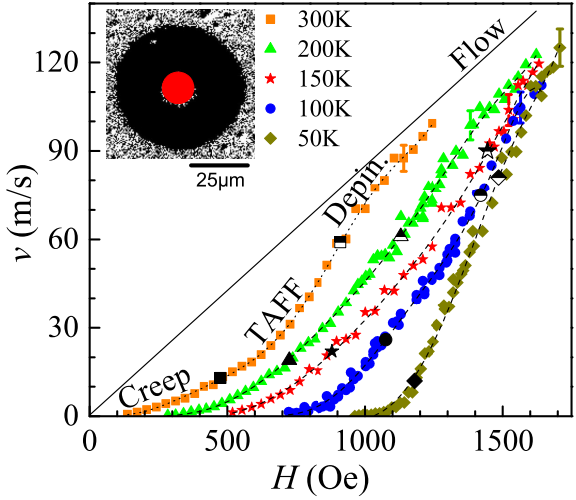}
\caption{\label{fig:raw_data}Variation of the domain wall velocity in the Pt/Co(0.45~nm)/Pt film with $H$ for different temperatures. The big black filled and half-filled symbols correspond to boundaries between the \textit{creep} and \textit{TAFF} regimes, $H_{C-T}$, and \textit{TAFF} and \textit{depinning} regimes, $H_{dep}$, respectively. The straight line represents the predicted asymptotic high field \textit{flow} limit at 300~K for $m =$ 0.085~m/s$\cdot$Oe. Inset: Domain wall displacement (in black) from a nucleus (in red) produced by a 1~$\mu$ s field pulse of amplitude $H=$865~Oe, at 150~K.}
\end{figure}

The experimental results were obtained with a sputter-grown ultrathin Pt(3.5~nm)/Co(0.45~nm)/Pt(4.5~nm) film deposited on an etched Si/SiO$_2$ substrate \cite{8}. The Curie temperature ($T_c =$375~K), square perpendicular hysteresis loop, thermal dependence of the saturated magnetization -- compatible with that predicted for an anisotropic 2D-ferromagnet -- and of the anisotropy field, were determined from magneto-optical Kerr magnetometry and microscopy measurements~\cite{sup_mat}. The film was cooled in an open cycle optical cryostat and its temperature was measured with an accuracy of $\pm$2~K. The DW motion was produced by field pulses and visualized by Kerr microscopy with a resolution of $\sim 1~\mu$m. The DW velocity, $v(H)$, was deduced from the growth in diameter of magnetically reversed bubbles (Fig.~\ref{fig:raw_data}, inset). The field pulses (0$ < H < $ 1600~Oe) were produced by a 60 turns small coil (0.16~$\mu$s rise time) positioned very close to the film surface\cite{
sup_mat}. For each
pulse amplitude and duration (from 0.5~$\mu$s to 100~$\mu$s), the DW displacements were analyzed in order to select only steady DW motion.

The DW velocity $v(H)$ curves are depicted in Fig.~\ref{fig:raw_data}, for temperatures ranging between 50 and 300~K. The main parameters controlling DW motion are reported in Table~\ref{tab:parameters}.
As it can be observed, lowering the temperature results in a shift towards the high field region of the curves. Different dynamical regimes when rising the applied magnetic field $H$ are indicated in Fig.~\ref{fig:raw_data} on the highest temperature (300~K) curve. Their precise identification is not straightforward and will be discussed in detail below.
\begin{table*}
\centering
\begin{ruledtabular}
\begin{tabular}{lrrrrr}
$T$ (K) & 300 & 200 & 150 & 100 & 50 \\
\hline
$M_s$ (erg/G.cm$^3$) & 800(40) &1120(50) & 1260(60) & 1370(70) & 1470(80)\\
$m$(m/s.Oe) & 0.085(0.04) & 0.083(0.05) & 0.091(0.05) & 0.089(0.06) & 0.102(0.07)\\
$H_{C-T}$(\textit{creep}) (Oe)  & 480(20) & 720(30) & 870(40) & 1080(40) & 1170(40)\\
$H_{C-T}$(\textit{TAFF}) (Oe)  & 470(20) & 730(30) & 890(40) & 1070(40) & 1190(40)\\
$H_{dep}$(\textit{TAFF}) (Oe)  & 920(30) & 1060(50) & 1360(30) & 1480(50) & 1500(40)\\
$H_{dep}$(\textit{depin.}) (Oe) &900(40) &  1190(40) & 1335(50) & 1360(50) & 1470(50)\\
$T_{dep}$ (K) & 1880(110) & 2500(150) & 2700(160) & 3200(190) & 3500(210)\\
\end{tabular}
\caption{\label{tab:parameters}Parameters controlling DW dynamics at different temperatures. $M_s$ is the magnetization saturation and $m$ the DW mobility in the \textit{flow} regime \cite{sup_mat}. Between the field boundaries $H_{C-T}$ and $H_{dep}$, DWs follow a thermally assisted flux flow (\textit{TAFF}) regime. The two sets of $H_{C-T}$ and $H_{dep}$ values, were determined by independent methods (see text) and are consistent together. The $T_{dep}$ value was deduced from the \textit{creep} regime data.}
\end{ruledtabular}
\end{table*}
The high field $v(H)$ response is supposed to finally reach the asymptotic \textit{flow} regime with a mobility $m = v/H$ (as shown in Fig.~\ref{fig:raw_data}) \cite{8}. Unfortunately, the determination of $v$ for $H >$1700~Oe was not possible due to the increase of the nucleation rate of magnetization reversal\cite{sup_mat}. The value of $m$ (reported in Table~\ref{tab:parameters}) is then determined from the expected power law expression~\cite{21}, $m-(v/H)=DH^{-c}$ with $c=4$. A small increase of $m$ is revealed when lowering the temperature~\cite{sup_mat}.
\begin{figure}
\includegraphics[width=.9\columnwidth]{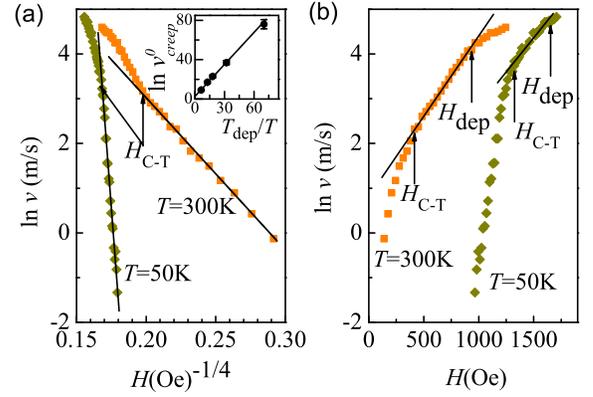}
\caption{\label{fig:TAFF}Field dependent domain wall velocity. (a) Plot of $\ln v$ vs $H^{-1/4}$ to evidence the \textit{creep} regime and its upper boundary $H_{C-T}$ (\textit{creep}). Inset: \textit{creep} pre-factor $v^0_{creep}$ deduced from an extrapolation of the \textit{creep} law (straight line) to $H^{-1/4}=$0. (b) Plot of $\ln v$ vs $H$ reveals the \textit{TAFF} regime and its two boundaries $H_{C-T}$ (\textit{TAFF}) and $H_{dep}$ (\textit{TAFF}).}
\end{figure}

Let us now identify the low field regimes in detail.
Two different behaviors for the velocity can be clearly seen in Fig.~\ref{fig:TAFF}. At lower fields $\ln v$ exhibits a linear variation with $H^{-1/4}$ and this part can be identified with the \textit{creep} regime, in
which the barriers diverge with the inverse force~\cite{feigelman,nattermann,1}.
In this regime, the DW velocity can be written:
\begin{eqnarray}
\label{eq:creep}
v_{creep}(H,T) =v^0_{creep}(T) e^{-\frac{T_{dep}}{T}\left(\frac{H_{dep}}{H}\right)^\mu} 
\end{eqnarray}
where $v^0_{creep}$ corresponds to a velocity prefactor whose meaning is discussed later, and $T_{dep}$ is the depinning temperature which is defined below. As observed in Fig.~\ref{fig:TAFF}(a), the \textit{creep} law is no longer fulfilled at high (low) $H$-values ($H^{-1/4}$-values) and a new regime of transport can be identified above the field $H_{C-T}$ \textit{(creep)},
given in Table~\ref{tab:parameters} and which corresponds to the upper boundary of the \textit{creep} regime.

For $H\geq H_{C-T}$ \textit{(creep)}, $\ln v$ varies linearly with $H$, as shown in Fig.~\ref{fig:TAFF}(b). The experiment thus reveals a regime in which the barriers decrease linearly as the field is increased and the velocity obeys
\begin{eqnarray}
\label{eq:TAFF}
v_{\textit{TAFF}}(H,T) = v^0_{\textit{TAFF}}(T) e^{-\frac{T_{dep}}{T}\left(1-\frac{H}{H_{dep}}\right)} 
\end{eqnarray}
where $v^0_{\textit{TAFF}}(T)$ is the velocity at $H=H_{dep}$. This regime can be identified with the so-called \textit{TAFF} regime~\cite{anderson} initially proposed in the context of vortex motion and  is compatible with the computed behavior in a $T=0^+$ dynamics~\cite{6}.
As shown in Fig.~\ref{fig:TAFF}(b), the limits of the \textit{TAFF}  regime ($\ln v \sim H$) permit to obtain $H_{C-T}$(\textit{TAFF}) and the depinning threshold, $H_{dep}$(\textit{TAFF}), reported in Table~\ref{tab:parameters}. More strictly, the end of the \textit{TAFF} regime corresponds to the field ($H\sim H_{dep}(1-T/T_{dep})<H_{dep}$) at which the barrier is of the order of the temperature, and at small temperature this field is expected to approach $H_{dep}$.

Beyond $H_{dep}$ the system enters in the depinning regime for which universal scaling forms of the field and temperature are expected  \cite{5,19,20,21,22}:
\begin{equation}
\label{eq:scaling}
v_{dep}(H,T) = v^0_{dep}(T) G\left[  \frac{H-H_{dep}}{H_{dep}}\left(\frac{T}{T_{dep}}\right)^{-\psi/\beta} \right]
\end{equation}
where $G(x)$ is a universal function such that
$G(x) \sim x^\beta$ for $x \gg 1$, i.e., $T \ll T_{dep}$, with $\beta$ the depinning exponent, and $v^0_{dep}=v_{dep}(H=H_{dep},T) =mH_{dep}(T/T_{dep})^\psi$, with $\psi$ the thermal rounding exponent. Thus, the velocity scaling in the depinning regime provides an alternative for estimating $H_{dep}$\cite{sup_mat}, according to the procedure proposed in Ref.~\cite{11}, by considering that the DW velocity fits a $v\sim(H-H_{dep} )^\beta$ law, where $\beta=$0.25~\cite{14}. This procedure leads to $H_{dep}(depin.)$ in Table~\ref{tab:parameters}. Knowing $H_{dep}$, the values for $T_{dep}$ reported in Table~\ref{tab:parameters} were deduced from the \textit{creep} plot slope, $T_{dep} H_{dep}^{1/4}/T$. Note that the different estimation methods lead to consistent values for $H_{C-T}$ and $H_{dep}$,with $H_{C-T}=$ 0.5-0.8 $H_{dep}$, as 
shown in Table~\ref{tab:parameters}.

The theoretical analysis of the data thus provides a consistent picture of \emph{all} the regimes. Their matching allows to extract additional results. We in particular analyze the temperature variation of the \textit{creep} velocity pre-factor $v^0_{creep}$ (cf. Eq.~(\ref{eq:creep})), shown in the inset of Fig.~\ref{fig:TAFF}(a).  The values of $\ln v^0_{creep}$ are deduced from the extrapolation to $H^{-1/4}=0$ of the \textit{creep} plot of Fig.~\ref{fig:TAFF}(a). Empirically, the velocity $v^0_{creep}$ is found to fit well with an exponential variation $v^0_{creep}=v'^0_{creep} \exp(C T_{dep}/T)$, over the explored temperature range $T/T_{dep}=$0.014 to 0.160. The best fit gives $C =$ 1.04$\pm$0.02 and $v'^0_{creep}=$35$\pm$15m/s. This strongly suggests that the velocity $v'^0_{creep}$ must be written as $v'^0_{creep}=\xi/\tau$, where $\xi$ is the disorder correlation length, and $1/\tau$ the attempt-frequency at passing barriers. Note that such a temperature dependent prefactor simply corresponds, 
in the \textit{creep} regime to a subdominant correction of the barrier. Extracting the explicit temperature dependence allows to obtain the microscopic attempt frequency. Since $\xi \approx$19nm (to be discussed below), one obtains a reasonable time scale of $\tau \approx $1ns. Using this exponential behavior of $v^0_{creep}$ and matching the \textit{creep} and \textit{TAFF} velocities at the crossover field $H_{C-T}$, a value of $C=$0.66$\pm$0.08 can be obtained, which is of the same order of magnitude and almost temperature independent, as the previously extracted $C$ value.

\begin{figure}
\includegraphics[width=.9\columnwidth]{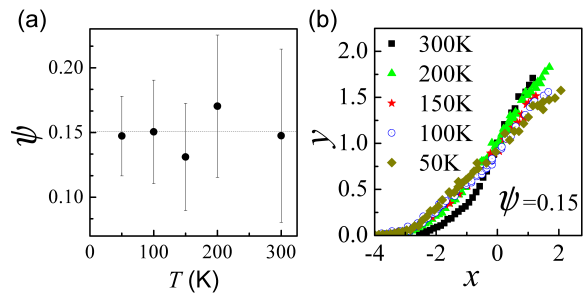}
\caption{\label{fig:scaling}(a) $\psi$ as a function of $T$ calculated from the crossover at $H=H_{dep}$ between the \textit{TAFF} and thermal rounding regimes, $v^0_{\textit{TAFF}}/(m H_{dep})= (T/T_{dep})^\psi$. (b) Universal scaling plot of the depinning transition using $\psi=0.15$, $\beta=$0.25 and the parameters of Table~\ref{tab:parameters}. The reduced coordinates are: $x=((H-H_{dep} )/H_{dep} ) (T/T_{dep} )^{-\psi/\beta}$and $y=(v/(mH_{dep} )) (T/T_{dep} )^{-\psi}$.
}
\end{figure}

The crossover between \textit{TAFF} and depinning is also consistent with the expected thermal rounding of the velocity in the depinning regime. Matching the \textit{TAFF} and the depinning regime at $H_{dep}$, the prefactor in Eq.~(\ref{eq:TAFF}) should be $v^0_{\textit{TAFF}} = v_{dep}(H=H_{dep},T)$, allowing to use the velocity corresponding to the upper bound of the \textit{TAFF} regime to extract the thermal rounding exponent $\psi$. As shown in Fig.~\ref{fig:scaling} the values are consistent with the prediction ($\psi=0.15$) of numerical simulations based on the Edwards-Wilkinson equation describing the overdamped motion of an elastic interface in a weak quenched disorder~\cite{5}.

In order to complete this analysis of universal exponents (i.e. the universality class) we now determine non-universal (intrinsic) characteristic length and energy scales controlling the pinning and their temperature dependence.
The free energy, $F(L, u)$, of a DW segment of length $L$, displaced over a distance $u$ under the action of the magnetic field writes:
\begin{equation}
\label{eq:free_energie}
F(L,u)=\epsilon_{el}u^2 / L -\epsilon_{pin}u\sqrt{n_i \Delta L}-M_s H t L u
\end{equation}
where $n_i$ is the density of pinning centers assumed to be equal to $1/\xi^2$. $M_s$ is the measured magnetization at saturation (Table~\ref{tab:parameters}), $t=$0.45~nm the Co layer thickness, and $\Delta =\sqrt{A/K_{eff}}$ the DW thickness parameter.  The first term in Eq.~(\ref{eq:free_energie}) represents the elastic energy of the wall, $F_{el}$, the second is the pinning energy, $F_{pin}$, while the third term stands for the Zeeman contribution, $F_H$. 
From this free energy we can derive expressions for the fundamental energy and length scales. In particular, the Larkin length, i.e. the length scale below which elasticity dominates over disorder, has direct experimental relevance~\cite{18}. The Larkin length,  $L_c$, supposed to be larger than $\xi$,  and the depinning field, $H_{dep}$, can be defined from the equalities $F_{el} (L_c,\xi)=F_{pin} (L_c,\xi)$ and $F_{el} (L_c,\xi)=F_H (L_c,\xi)$.

Assuming that $k_B T_{dep}=F_{pin} (L_c,\xi) (=\epsilon_{pin} \sqrt{\Delta L_c})$, the following expressions hold:
\begin{eqnarray}
\label{eq:relation1}
\epsilon_{el}&=&(k_B T_{dep})^2 / (M_s H_{dep} t)\xi^3 \\
\label{eq:relation2}
\epsilon_{pin}^2&=&[(k_B T_{dep})(M_s H_{dep} t) \xi]/\Delta \\
\label{eq:relation3}
L_c&=&(k_B T_{dep}) / [(M_s H_{dep} t) \xi] 
\end{eqnarray}
The DW elastic energy density expresses as $\epsilon_{el}=4t\sqrt{AK_{eff}}$. The estimated values of the exchange stiffness $A$ and the effective anisotropy $K_{eff}$ for the $t=$0.45~nm thick film were deduced from data obtained with thicker films~\cite{sup_mat}. For $T=300~K$, we get $A=$1.25~$\mu$erg/cm, an anisotropy field
$H_{eff}^{A}=$5.3$\pm$0.3~kOe, and $K_{eff}=(H_{eff}^{A} M_s) /2 =$2.1$\pm$0.3~Merg/cm$^3$, which leads to $\epsilon_{el}=$0.29$\pm$0.04~$\mu$erg/cm. Starting from Eq.~(\ref{eq:relation1}), and accounting for the values of $H_{dep}$ and $T_{dep}$, reported in Table~\ref{tab:parameters}, we estimate $\xi=$19.3~nm. This value agrees well with the mean lateral size of Pt crystallites ($\sim $10-20~nm) determined by AFM \cite{24} in such Pt/Co/Pt film supposing a morphological continuity between Pt and Co layers. Therefore we assume that $\xi$ is temperature independent in the following analysis.
\begin{figure}
\includegraphics[width=.9\columnwidth]{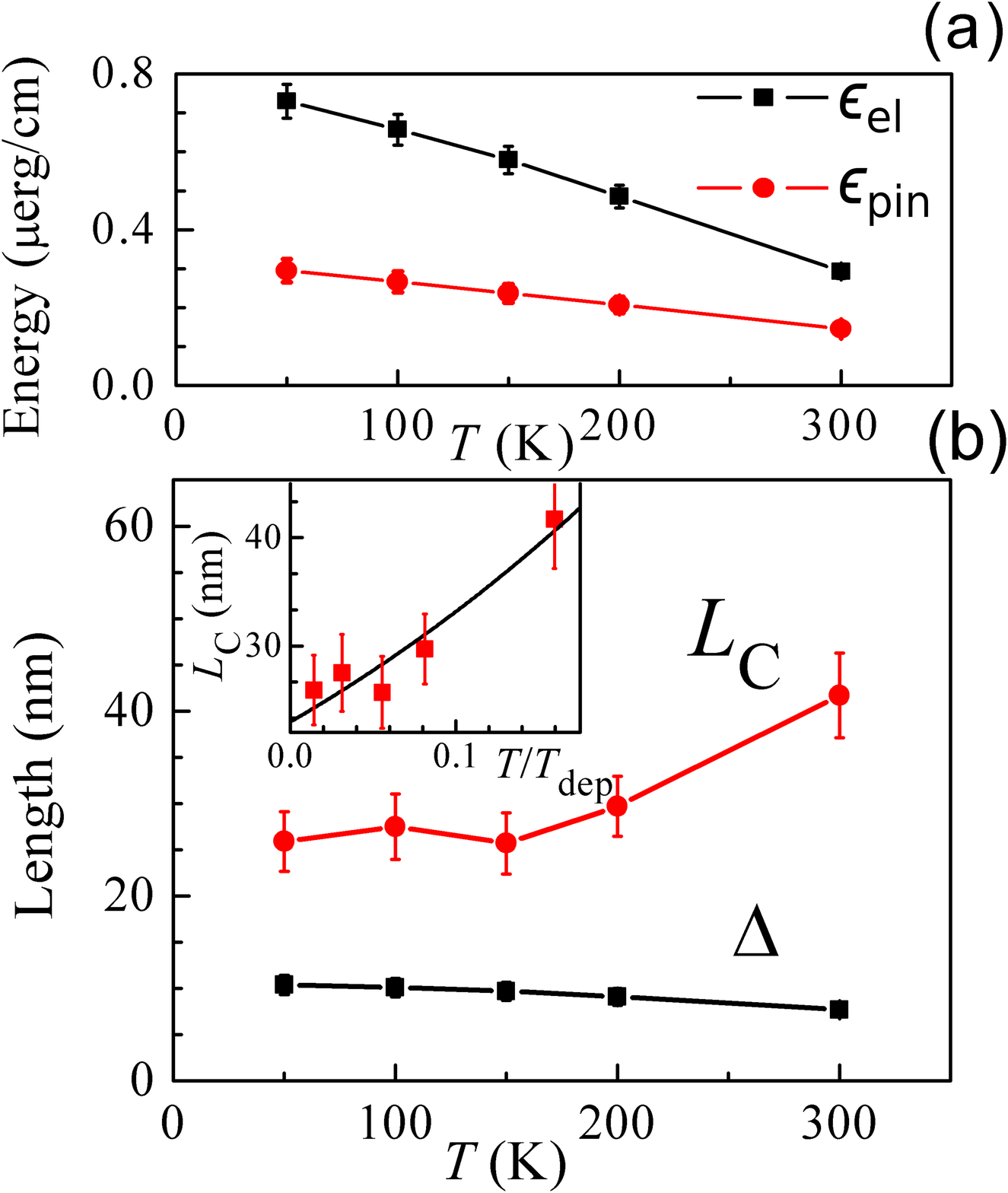}
\caption{\label{fig:energies}: (a) Temperature variation of the elastic energy density, $\epsilon_{el}$, and the pinning energy density, $\epsilon_{pin}$. We used the value of $\xi$($=$19.3~nm), obtained for $T=$300~K. (b) Temperature dependence of the wall width parameter, $\Delta$, and of the Larkin length, $L_c$. In the inset, $L_c$ is plotted versus $T/T_{dep}$ . The line is the fit of Eq.~(\ref{eq:larkin_length}) used for determining the $\phi$ exponent.}
\end{figure}
We can then estimate the temperature dependence of $\epsilon_{el}$, $\epsilon_{pin}$, $\Delta$, and $L_c$. $H_{eff}^{A}(T)$ is found to be nearly constant over the 50-300~K temperature range, which means that $K_{eff} (T)\sim M_s (T)$. For $A(T)$, we assumed a temperature variation $\sim M_s (T)^2$. As a result, $\epsilon_{el}$ varies as $M_s (T)^{3/2}$ and $\Delta$ as $M_s (T)^{1/2}$. This more pronounced temperature variation for $\epsilon_{el}$ than for $\Delta$ is compatible with the results reported in Figs.~\ref{fig:energies}(a) and \ref{fig:energies}(b). As expected, $\Delta$ decreases when rising the temperature consistently with the found thermal variation of the flow mobility $m$ (Table~\ref{tab:parameters}). Alternatively, $\epsilon_{el} (T)$ was deduced from Eq.~(\ref{eq:relation1}), using the values of $M_s$, $H_{dep}$ and $T_{dep}$ reported in Table~\ref{tab:parameters}. As shown in Fig. \ref{fig:energies}, $\epsilon_{el}$ decreases as temperature increases. The variation is however 
weaker than 
expected,
probably because of the crude assumption: $\Delta \ll \xi$, while both values (7.7~nm and 19.3~nm, respectively) are quite close. The density of pinning energy $\epsilon_{pin}$ is deduced from Eq.~(\ref{eq:relation2}). Fig.~\ref{fig:energies} shows that $\epsilon_{pin}$ exhibits a weaker temperature variation than $\epsilon_{el}$. 

Finally, the temperature dependence of the Larkin length, $L_c$, is calculated from Eq.~(\ref{eq:relation3}) and the results shown in Fig.~\ref{fig:energies}(b). $L_c (T)$ can be recast in the form \cite{23,25,26}:
\begin{equation}
\label{eq:larkin_length}
L_c(T)=L_c^0[1+(T/T_{dep})]^{1/\phi}
\end{equation}
where $L_c^0= L_c (T=0K)$ and $\phi$  is the thermal crossover exponent of the Larkin length. Fitting $L_c (T)$ with Eq.~(\ref{eq:larkin_length}) (see the inset of Fig.~\ref{fig:energies}(b)) leads to $L_c^0=$25$\pm$2~nm and $\phi=$0.24$\pm$0.05. This last value agrees with the most recent prediction, $\phi=$1/5 \cite{23,25,26}.

In conclusion, our consistent analysis of the thermal dependence of the field-driven DW motion would be valuably extended to pinning mechanism studies in other systems.

{\it - Acknowledgements:}
We wish to thank A. Mougin and P. Metaxas for fruitful discussions and B. Rodmacq for providing us high quality films. S.B., J.G. and V. J. acknowledge support by the French-Argentina project ECOS-Sud num.A12E03.  This work was also partly supported by the french projects DIM C’Nano IdF (Region Ile-de-France). S.B. and A.B.K. acknowledge partial support from Project PIP11220090100051 (CONICET). This work was supported in part by the Swiss NSF under MaNEP and Division II.


\end{document}